\documentclass[aps,prd,twoside,tightenlines,superscriptaddress,reprint,showpacs,preprintnumbers,nofootinbib,floatfix]{revtex4-1}
\usepackage{graphicx}
\usepackage[sort&compress]{natbib}
\usepackage{amsmath,amssymb,amsfonts}
\usepackage[T1]{fontenc}
\usepackage{lmodern,dsfont}
\usepackage[colorlinks=true,urlcolor=blue,linktoc=page,linkcolor=blue,citecolor=blue]{hyperref}
\usepackage{booktabs}
\newcommand{\Ima}{\textrm{Im}}

\newcommand{\mev}{\textrm{ MeV}}
\newcommand{\be}{\begin{equation}}
\newcommand{\ee}{\end{equation}}
\newcommand{\ba}{\begin{eqnarray}}
\newcommand{\ea}{\end{eqnarray}}

\newcommand{\nn}{{\nonumber}}

\begin{document}

\title{\boldmath Finite volume treatment of $\pi\pi$ scattering in the $\rho$ channel}
\date{\today}

\author{M.~Albaladejo}
\affiliation{Instituto de F\'isica Corpuscular (centro mixto CSIC-UV), Institutos de Investigaci\'on de Paterna, Aptdo.~22085, 46071, Valencia, Spain}
\author{G.~Rios}
\affiliation{Helmholtz-Institut f\"ur Strahlen- und Kernphysik, Universit\"at Bonn, D-53115 Bonn, Germany}
\author{J.~A.~Oller}
\affiliation{Departamento de F\'{\i}sica. Universidad de Murcia. E-30071 Murcia, Spain}
\author{L.~Roca}
\affiliation{Departamento de F\'{\i}sica. Universidad de Murcia. E-30071 Murcia, Spain}


\begin{abstract}
We make a theoretical study of $\pi\pi$ scattering  with quantum numbers $J^{PC}=1^{--}$ 
in a finite box. To calculate physical observables for infinite volume from lattice QCD, 
the finite box dependence of the
potentials is not usually considered.  We quantify such effects by means of two different approaches for
vector-isovector  $\pi\pi$ scattering based on Unitarized Chiral Perturbation Theory results: the  Inverse Amplitude Method and another one based on the $N/D$ method. We take into account finite box effects stemming from
 higher orders through loops in
the crossed $t,u-$channels as well as from the renormalization of the coupling
constants. The main conclusion is that for $\pi\pi$ phase shifts in the isovector
channel  one can safely apply L\"uscher based methods for finite box sizes of
$L$ greater than $2 m_\pi^{-1}$. 

\end{abstract}

\maketitle

\section{Introduction}
\label{Intro}

Along years significant efforts have been devoted to 
determining the hadron spectrum with lattice QCD calculations
\cite{Nakahara:1999vy,Mathur:2006bs,Basak:2007kj,Durr:2008zz,Bulava:2010yg,Morningstar:2010ae,Foley:2010te,Alford:2000mm,Kunihiro:2003yj,Suganuma:2005ds,Hart:2006ps,Wada:2007cp,Prelovsek:2010gm,Lin:2008pr,Gattringer:2008vj,Engel:2010my,Mahbub:2010me,Edwards:2011jj,Lang:2011mn,Prelovsek:2011im}.
We refer to Ref.~\cite{reviewhadron} for a recent  review on the different
methods and results. A crucial feature of the lattice QCD studies
is that one must extrapolate from the finite box used in the lattice
calculations to 
infinite space in order to obtain physical
observables. To this aim, one of the tools most widely used is the
L\"uscher's approach \cite{luscher,Luscher:1990ux}. At the practical level, the latter 
method was recently rederived in a simpler way in Ref.\cite{misha} by employing 
general expressions for two-body scattering from Ref.~\cite{Oller:1998zr}. It is 
worth stressing that the method of Ref.~\cite{misha} keeps the relativistically covariant 
two-body propagator. This method has been applied in
Refs.~\cite{mishajuelich, MartinezTorres:2011pr, Roca:2012rx, mishakappa, Albaladejo:2012jr, Albaladejo:2013aka, MartinezTorres:2011pr} 
 to calculate physical observables from possible calculations of different hadronic processes in a finite box as well as associated errors.

However, a common feature of Refs.~\cite{luscher,Luscher:1990ux,misha}
 is that  they do not take into account the finite size dependence of 
potentials\footnote{The potentials are analytical functions without the elastic two-body unitarity cut.}
 from which the scattering in infinite volume is obtained.
This simplification is based on the result of Ref.~\cite{luscher} that the
  volume dependence of the potentials is  ``exponentially
suppressed'' with the box volume. 
 In Ref.~\cite{Albaladejo:2012jr} the effects of the finite size in 
 the isoscalar and isotensor $\pi\pi$ $S$-wave  
scattering   channels were studied quantitatively. 
 An estimation was given of the minimum box size of possible lattice calculations so that the errors coming from these finite size effects were negligible.

Following a similar procedure as in Ref.~\cite{Albaladejo:2012jr}, the
main aim of the present work is to analyze the effects of finite volume
for  $1^{--}$ $\pi\pi$  scattering in order to extract
physical observables for infinite volume from lattice QCD. This channel is of special 
relevance since it has the quantum numbers of the resonance $\rho(770)$  whose study 
 in  lattice QCD has attracted special
attention in the last decade. Several works have used L\"uscher's method to
extract physical observables in infinite volume from actual lattice QCD
data \cite{Durr:2008zz, Feng:2011ah, Aoki:2007rd, Gockeler:2008kc, Aoki:2010hn, Feng:2010es, Frison:2010ws,Dudek:2012xn}. 
The method of Ref.~\cite{misha} was applied in Ref.~\cite{Chen:2012rp}
in order to theoretically illustrate 
 the improvement of using the method of 
Ref.~\cite{misha} for determining  $\pi\pi$
scattering in the $\rho$ channel from  finite box calculations 
over the standard L\"uscher's method. In
Ref.~\cite{Chen:2012rp} the potential for the $\pi\pi$ scattering in
$I=1$ and $P$-wave was obtained from Ref.~\cite{Oller:2000ug} which
considered as dynamical input  
tree level diagrams from lowest order
Chiral Perturbation Theory (ChPT)
 with explicit exchange of vector resonances.
In Ref.~\cite{Chen:2012rp}  the potential is not modified in the finite box and,
hence,  the finite volume effects come only
from the discretization in the box of the unitarization loops. In the
present work we use the amplitudes from the Inverse Amplitude Method (IAM) 
and another method based on an approximate algebraic solution to the N/D method that 
we refer as the $N/D_A$ in the following. In both cases the amplitudes include
 loops in the $t-$ and $u-$ channels,
which indeed get modified in the finite volume. This provides two
different volume dependent amplitudes for the isovector $\pi\pi$ $P$-wave scattering. 
We can then study the impact of 
 volume dependence from these amplitudes in the process of 
extracting  physical observables in the infinite volume, like
phase shifts, from scattering energy levels
obtained in the finite box. The error made by neglecting possible
volume dependence of the potential in the direct L\"uscher's analysis of the
data is discussed.


\section{Formalism}
For the study of the finite volume effects in $\pi\pi$ interactions in the $\rho$ channel, we will consider two approaches, the $N/D_A$ method and the IAM, whose dynamical input is obtained from
 the $SU(2)$ chiral amplitudes at $\mathcal{O}(p^4)$.\footnote{We consider elastic $\pi\pi$ amplitudes, since the $K\bar{K}$ threshold is not very relevant for the $\rho$ resonance \cite{Oller:1998zr}. We do not consider either the $4\pi$ channel contribution  discussed in Ref.~\cite{Albaladejo:2012jr}, following the findings of Ref.~\cite{Albaladejo:2008qa}.} These chiral amplitudes are discussed in Subsec.~\ref{subsec:amplitudes_p4}, whereas the amplitudes involving $\rho$ exchanges are considered 
 in Subsec.~\ref{subsec:RhoAmplitudes}. In  finite volume, since the loop integrals are replaced with sums over discrete momenta, these amplitudes are volume dependent. The finite volume effects on the amplitudes are treated in Subsec.~\ref{subsec:fin_vol}. In Subsecs.~\ref{subsec:ND} and \ref{subsec:IAM} we discuss the $N/D_A$ and the IAM methods, respectively, and show how they make use of the one-loop perturbative chiral amplitudes discussed before. The amplitudes depend on some free parameters, that are fixed in Subsec.~\ref{subsec:fits} by reproducing several sets of data.

\subsection{\boldmath Chiral amplitudes at $\mathcal{O}(p^4)$}
\label{subsec:amplitudes_p4}

Let us denote the different $\pi\pi$ partial waves of isospin $I$ and angular momentum $J$ by $T^{IJ}(s)$. These amplitudes are normalized such that:
\begin{equation}
T^{IJ}(s) = -\frac{8\pi\sqrt{s}}{p} \frac{1}{\cot \delta^{IJ}(s) - i}~,\label{eq:defTmatphase}
\end{equation}
where $p = \sqrt{s/4 - m_\pi^2}$ is the CM momentum of each pion (with mass $m_\pi$), and $\delta^{IJ}(s)$ is the partial wave phase shift. These amplitudes can be calculated perturbatively as the projection into angular momentum $J$ wave of the different isospin amplitudes $A^I(s,t,u)$, that are computed in $SU(2)$ ChPT \cite{Gasser:1983yg}. Here, $s$, $t$ and $u$ are the usual Mandelstam variables. We denote by $A_{2n}^I(s,t,u)$ the $\mathcal{O}(p^{2n})$ contribution to this amplitude, and by $T^{IJ}_{2n}(s)$ its projection into 
the partial wave with angular momentum $J$. Both the IAM and the $N/D_A$ methods, as we shall see below, make use of the amplitudes $T^{IJ}_{2n}(s)$ to calculate the partial waves $T^{IJ}(s)$ in a non-perturbative manner, in the sense that they resum the unitarity cut. (In what follows, the superscript $IJ$ is dropped to simplify  notation.)

The amplitude $A_4$ can be written in a generic way, for the different isospin channels, as:
\begin{align}
A_4(s,t,u) = & P_L + P_H H(m^2) +\nn \\
 & P_{G,s}G(s) + P_{G,t} G(t) + P_{G,u} G(u)~. \label{A_4}
\end{align}
The functions $P_X$ above are polynomials of the Mandelstam variables. In particular, the low energy constants (LECs) $\bar{l}_i$ appear just in the term $P_L$.\footnote{We work here with the finite and scale-independent LECs, $\bar{l}_i$. In the case of $\pi\pi$ scattering, only the LECs with $i=1,\ldots,4$ are involved.} In Eq.~\eqref{A_4}, $H$ and $G(P^2)$ are the one-- and two--point one--loop functions, respectively, given by:
\begin{align}
  G(P^2) & = \int\frac{d^3\vec q}{(2\pi)^3} I(\vec{q},P)~, \label{G_de_P}\\
  I(\vec{q},P) & =
    \frac{(\omega_{\vec q}+\omega_{\vec P-\vec q})/(2\omega_{\vec q}\,\omega_{\vec P-\vec q})}{(P^0-\omega_{\vec q}-\omega_{\vec P-\vec q})
      (P^0+\omega_{\vec q}+\omega_{\vec P-\vec q})}~, \label{eq:I(q,P)}\\
  H = & \int\frac{d^3\vec q}{(2\pi)^3}\frac1{2\omega_{\vec q}}~,\label{tadpole}
\end{align}
where $P$ is the four--momentum entering the loop. Whence,
$G(s)$, $G(t)$ and $G(u)$ in Eq.~\eqref{A_4} stem from the
$s$--, $t$-- and $u$--channel loops~\eqref{G_de_P}
with $P^2=s$, $t$ and $u$, respectively. These functions are regularized through dimensional regularization. After the divergences and scale dependencies are absorbed in the LECs~\cite{Gasser:1983yg}, the loop function, denoted now $G^D$, reads:
\begin{equation}
  \label{eq:loopfun}
  G^D(P^2) = \frac{1}{16\pi^2} \left( -1 + \sigma(P^2) \mathrm{log} \frac{1+\sigma(P^2)}{1-\sigma(P^2)} \right)~,
\end{equation}
with $\sigma(P^2) = \sqrt{1-4m_\pi^2/P^2}$. On the other hand, because of the regularization approach followed, we have $H^D = 0$.

\subsection{\boldmath Chiral amplitudes and the $\rho$ meson}
\label{subsec:RhoAmplitudes}
From a broader point of view, one can include, together with the $\mathcal{O}(p^4)$ amplitudes explained so far, the contribution coming from Resonance ChPT \cite{Ecker:1988te, Ecker:1989yg}, which start to contribute at one-loop order as well. 
 For the  problem discussed here, 
 the relevant Lagrangian is the one that incorporates exchanges of the $\rho$ meson field. This term involves, in our case, two  free parameters, the bare $\rho$ mass, $M_\rho$, and the  coupling constant $G_V$. This contribution is relevant in the $N/D_A$ method for the $I=1$, $J=1$ channel, since the physical $\rho$ resonance is generated by dressing the bare $\rho$ pole with $\pi\pi$ rescattering \cite{Oller:1998zr}. We also calculate the $\rho$ exchanges in the $t$-- and $u$--channels for 
 the $I=0,~2$ $\pi\pi$ $S$-wave channels which are also studied here for consistency. These exchanges cancel partially with the  crossed channel $\pi\pi$ loops \cite{Oller:1998zr}. 

It is well known that the exchange of resonances has a large impact on the values of the LECs. The contribution of the $\rho$ resonance to these constants reads \cite{Ecker:1988te}:
\begin{eqnarray}
\bar{l}_1^{\rho} & = - 96\pi^2 \frac{G_V^2}{M_\rho^2}~, \nn \\
\bar{l}_2^{\rho} & = \hphantom{+} 48\pi^2 \frac{G_V^2}{M_\rho^2}~, \\
\bar{l}_3^{\rho} & = \bar{l}_4^{\rho} = 0~. \nn
\end{eqnarray}
Since these contributions are already accounted for by the $\rho$-exchange amplitudes, we subtract them from the LECs appearing in the amplitudes to avoid double counting. 
 This is the case when applying the unitarization scheme based on the N/D method, but  not in the IAM. In this latter method the physical $\rho$ is generated from the dynamics through the ChPT ${\cal O}(p^4)$ LECs $\bar{l}_i$ \cite{Dobado:1996ps}. This makes a difference between the $N/D_A$ method and the IAM, that will show up in the different finite volume effects predicted by each approach. This point is not a shortcoming of our study, but rather it will allow us to confront both methods, so that the differences between both are considered as an estimate for uncertainties. 

\subsection{Finite volume effects}
\label{subsec:fin_vol}
Let us now discuss the modifications to the amplitudes explained so far when one considers the interactions in a finite cubic box of edge $L$ and with periodic boundary conditions. The chiral amplitude $A_4(s,t,u)$ receives contribution from loop integrals that, in the finite volume case, are replaced with sums over the allowed (quantized) momenta, giving rise to $\widetilde{A}_4(s,t,u)$, which is the finite volume version of the former. Namely, these contribution arise from $s$--, $t$-- and $u$--channel loops, 
 and from tadpole diagrams as well. Note also that we write the amplitudes in terms of the physical pion mass $m_\pi$ and decay constant $f_\pi$, and hence the $\mathcal{O}(p^4)$ contributions  to them (tadpole loop-functions) are included as $\mathcal{O}(p^4)$ terms in the amplitudes $A_4$ and $\widetilde{A}_4$. The $\mathcal{O}(p^4)$ contribution to the partial wave in the finite volume, that we denote by $\widetilde{T}_4$, is calculated as $T_4(s)$ with the replacement $A_4 \to \widetilde{A}_4(s,t,u)$. This one is calculated from Eq.~\eqref{A_4}, but replacing
the loop functions in Eqs.~\eqref{G_de_P} and \eqref{tadpole} 
with their finite volume counterparts,
$\widetilde G^D$ and $\widetilde H^D$. 
 Following the procedure in Ref.~\cite{MartinezTorres:2011pr}, the finite volume loop functions are obtained from the infinite volume ones as: 
\begin{align}
  & \widetilde G^D(P) =G^D(P^2)+ \nn\\
&   \lim_{q_\textrm{max}\to\infty}\Bigg[
  \frac1{L^3}\sum_{\vec{q}_i}^{q_\textrm{max}}I(\vec{q}_i,P)
   -\int\limits_{q<q_\textrm{max}}\frac{d^3\vec q}{(2\pi)^3}I(\vec{q},P)\Bigg]~,\label{eq:loopfun_fin}  \\
&  \widetilde H^D = H^D + \nn \\
& \lim_{q_\textrm{max}\to\infty}\Bigg[
  \frac1{L^3}\sum_{\vec{q}_i}^{q_\textrm{max}} \frac1{2\omega_{\vec{q}_i}}
  -\int\limits_{q<q_\textrm{max}}\frac{d^3\vec
  q}{(2\pi)^3}\frac1{2\omega_{\vec q}}\Bigg],\\
&  \vec{q}_i = \frac{2\pi}{L} \vec{n}~, \quad \vec{n} \in \mathbb{Z}^3~, \qquad \omega_{\vec{q}} = \sqrt{\vec{q}^2 + m_\pi^2}~, \nn
\end{align}
being $I(\vec{q},P)$ the integrand of Eq.~\eqref{G_de_P}, defined in Eq.~\eqref{eq:I(q,P)}. Since the box breaks Lorentz symmetry, the reference frame is fixed to
the center of mass frame of the initial pions. For this reason we have
used $P$ as the argument of $\widetilde G^D$ in Eq.~\eqref{eq:loopfun_fin} instead of $P^2$. For the $s$--channel loop case, where $\vec P=0$ so 
that $(P^0)^2=P^2=s$, we obtain $\widetilde G^D(P)$ as:
\begin{align}
& \widetilde G^D(s) = G^{D}(s) + \nn \\
& \lim_{q_\textrm{max}\to \infty}
\Bigg[ \frac{1}{L^3}\sum_{\vec{q}_i}^{q_\textrm{max}}I(\vec{q}_i,s)
- \int\limits_{q<q_\textrm{max}}\frac{d^3\vec{q}}{(2\pi)^3} I(\vec{q},s)
\Bigg]~,\label{tonediff}
\end{align}
and:
\begin{equation}
I(\vec{q},s)=\frac{1}{\omega(\vec q)}\,\,\frac{1}{s-4\omega(\vec q)^2}~.\label{prop_contado}
\end{equation}
Let us note that $\widetilde G^D(P)$ depends solely on $P^2=s$ in this case. However, for the $t$--channel loop, we have $P^0=0$ so that $P^2=-\vec P^2=t$, and hence the integrand $I(\vec{q},P)$ becomes:
\begin{equation}
  \label{eq:IqP_t}
  I(\vec{q},P)=-\frac{1}{2\omega_{\vec q}\,\omega_{\vec P-\vec q}\,(\omega_{\vec q}+
    \omega_{\vec P-\vec q})}.
\end{equation}
Now, contrary to the $s$--channel case, $G(P)$ depends on
$P^2=t$, but also on $\vec P$ and its relative orientation respect to
the cubic lattice of allowed momenta in the box, $\{\vec q_i\}$.
This dependence translates into a dependence on the scattering angle $\theta$,
present in $t=-2(s/4-m_\pi^2)(1-\cos\theta)$, but also on the
azimuthal angle $\phi$, and this also happens with the $u$--channel case.
Thus, when making the partial wave projections in finite volume,\footnote{The partial wave projection 
 in the infinite volume case restricts to integrate on $\cos\theta$,  
$T^{IJ}(s)=\frac1{2}\int d(\cos\theta)A^I(s,\cos\theta) P_J(\cos\theta)$~.} we should
now also integrate on $\phi$,
\begin{align} 
\widetilde{T}^{IJ}(s)=\frac1{4\pi}\int d\phi\int d(\cos\theta)A^I(s,\cos\theta,\phi) P_J(\cos\theta)~,
\end{align}
 being $P_J(x)$ the Legendre polynomial of order $J$. 

Finally, the function $\widetilde H^D$ can be computed using the Poisson resummation formula (see \textit{e.g.} Ref.~\cite{Bedaque:2006yi})
 and, since $H^D=0$, we find:
\begin{equation}
  \label{eq:Hgorro}
  \widetilde H^D = \frac{m_\pi}{4\pi^2L}\sum_{0\ne\vec n\in\mathbb Z^3}
  \frac1{|\vec n|}K_1(|\vec n|m_\pi L),
\end{equation}
being $K_1$ the Bessel function.


The $s$--channel loops are responsible for the right--hand cut (RHC), or unitarity cut, 
that stems from the loop function $G(s)$ and it is the most important source of $L$ dependence 
in the amplitude. This $L$ dependence arising from the RHC is the one used by the 
L\"uscher method to obtain phase shifts from the energy 
levels obtained in a finite volume. 
In particular, in the version of Ref.~\cite{misha}, given a certain energy 
level $E$ 
on a finite box of size $L$, the infinite volume amplitude is given by
\begin{align}
  \label{eq:euluscher}
  T^{-1}(E) & = \widetilde{G}^D(E) - G^D(E),
\end{align}
from where the phase shifts can be straightforwardly extracted.
Note that the $T$-matrix so calculated does not depend on any 
regularization procedure for the loop function $G$ since its inverse is  the 
limit of the piece between brackets in  Eq.~\eqref{eq:loopfun_fin}
(for further discussion on Eq.~\eqref{eq:euluscher}, see Ref.~\cite{Oset:2012bf}). 
Eq.~\eqref{eq:euluscher} only takes into account the $L$--dependence coming
from the $s$--channel loops. However, the tadpoles and the $t$-- and $u$--channel loops, which give rise to the left-hand cut (LHC) when projected into partial waves, give further dependence on $L$ (polarization corrections in the terminology of Ref.~\cite{Luscher:1990ux}) that is neglected in Eq.~\eqref{eq:euluscher}. Its effects are typically disregarded because they are exponentially suppressed \cite{Luscher:1990ux}. 

\subsection{\boldmath The $N/D_A$ method}
\label{subsec:ND}
From the exact $N/D$ method \cite{Chew:1960iv} it was shown in Ref.~\cite{Oller:1998zr} that 
 when the LHC is neglected or treated perturbatively (see also Refs.~\cite{Oller:1998ia, Oller:2000fj, Oller:2000ma,Albaladejo:2011bu, Albaladejo:2012sa}), a two-body partial wave amplitude can be written in full generality as:
\begin{equation}\label{Tinv}
T^{-1}(s) = V^{-1}(s) - G^S(s)~,
\end{equation}
where the loop function $G^S(s)$ is given by a once-subtracted dispersion relation of Eq.~\eqref{G_de_P},
\begin{equation}  \label{eq:Goncesub}
  G^S = \frac{1}{16\pi^2} \left( a(\mu) + \log \frac{m_\pi^2}{\mu^2} + \sigma(s) \mathrm{log} \frac{1+\sigma(s)}{1-\sigma(s)} \right)~.
\end{equation}
Here, $a(\mu)$ is a subtraction constant, that depends on the renormalization scale $\mu$, taken as $\mu = 770\ \text{MeV}$. The kernel $V$ is calculated by matching its chiral expansion, $V(s) = V_2(s) + V_4(s) + \cdots$, with the chiral amplitudes $T_2$ and $T_4$, that is:
\begin{align}
T(s) &= \frac{V(s)}{1 - V(s)G^S(s)} = V_2 + V_4 + V_2^2 G^S+\cdots \nn\\
& = T_2 + T_4 +\cdots  \label{eq:chiralexpansion}~,
\end{align}
where the ellipsis indicate ${\cal O}(p^6)$ and higher orders in the expansion. It results then:
\begin{align}
V_2(s) & = T_2(s)~, \nn \\
V_4(s) & = T_4(s) - T_2(s)^2 G^S(s)~. \label{kernels_4}
\end{align}
Note that the kernel $V_4$ has no RHC, since $T_4(s)$ contains the piece $T_2^2 G^D$ and then this
 cut cancels in the difference in Eq.~\eqref{kernels_4}. Hence, the RHC stems solely from the denominator in Eq.~\eqref{eq:chiralexpansion}. Now, to calculate the amplitude in the finite volume, $\widetilde{T}$, we must replace $V$ with $\widetilde{V} = V_2 + \widetilde{V}_4$. Since the $\mathcal{O}(p^2)$ term does not depend on $L$, no replacement is needed on it, whereas we have:
\begin{equation}
V_4 \to \widetilde{V}_4 = \widetilde{T}_4 - T_2 \widetilde{G}^S~,
\end{equation}
with $\widetilde{G}^S$ given by an analogous expression to Eq.~\eqref{tonediff} but with $G^D$ replaced 
by $G^S$. Notice that, according to the discussion above, the volume dependence of the $s$-channel loop does not affect the kernel $\widetilde{V}_4$. Hence, the volume dependence in the amplitude $\widetilde{T}$ enters through: (i) the kernel $V_4$ which volume dependence originates from tadpoles and  $t$-- and $u$-- loop functions generating   LHC, and (ii)  denominator $1-\widetilde{V}\widetilde{G}^S$, that gives rise to the RHC, through the function $\widetilde{G}^S$. This is the most important source of volume dependence. Finally, the energy levels $E$ within a box of edges of size $L$ are given by the poles of the amplitude $\widetilde{T}(s)$, $s=E^2$, that corresponds to the solution of the equation:
\begin{equation}
1 - \widetilde{V}\widetilde{G} = 0~.\label{eq:polosND}
\end{equation}
One can then reobtain the infinite volume phase shifts from the finite volume
energy levels obtained by using Eq.~\eqref{eq:euluscher} and quantify
the effect of neglecting the $L$--dependence in the kernel $V_4$.


\subsection{The Inverse Amplitude Method (IAM)}
\label{subsec:IAM}
Now, we consider the elastic IAM \cite{Dobado:1996ps, Dobado:1992ha, Dobado:1989qm, Truong:1991gv, Truong:1988zp}. This method uses elastic unitarity and ChPT \cite{Gasser:1983yg} to calculate the inverse of the $\pi\pi$ scattering partial wave $T^{IJ}$ through a dispersion relation.

In the IAM, one considers an auxiliary function, $W(s) \equiv T_2(s)^2/T(s)$, its analytic structure being a RHC from $4m_\pi^2$ to $\infty$, a LHC from $-\infty$ to $0$, and possible poles coming from zeros of the $T$ function (like Adler zeros). Hence, we can write a dispersion relation for $W$ as:
\begin{align}
W(s) & =W(0)+W'(0)s+\tfrac1{2}W''(0)s^2+ \nn\\
&  \frac{s^3}{\pi}\int_{\text{RHC}}ds'\frac{\Ima\, W(s')}{s'^3(s'-s)}+LC(W)+PC~.\label{eq:disp-rel}
\end{align}
The RHC integral can be evaluated exactly due to unitarity, whereas the subtraction constants can be evaluated with ChPT, since they involve amplitudes or derivatives of them at $s=0$, and so $W(0)\simeq T_2(0)-T_4(0)$, $W'(0)\simeq T_2'(0)-T_4'(0)$, $W''(0)\simeq -T_4''(0)$. The LHC is dominated by the low energy region, due to the three subtractions,
and it is also  dumped by an extra $1/(s'-s)$ for physical values of $s$. Then, it can be evaluated using ChPT to obtain $LC(W)\simeq-LC(T_4)$.
The pole contribution, $PC$, is formally ${\cal O}(p^6)$ and we neglect it 
(this causes some technical problems in the subthreshold region around the
Adler zeros which can be easily solved, but they do not
affect the description of scattering nor resonances, for details see Ref.~\cite{GomezNicola:2007qj}). Taking into account all the above considerations we arrive at the simple IAM formula, 
\begin{equation}
  \label{eq:IAM}
  T = \frac{T_2^2}{T_2-T_4}~.
\end{equation}
Hence, the amplitude in the box is obtained as $\widetilde{T}$ by replacing in Eq.~\eqref{eq:IAM} the amplitude $T_4$ with $\widetilde{T}_4$. The energy levels $E$ are then given by the poles of the amplitude $\widetilde{T}$, that is, by the solution of the equation $T_2 - \widetilde{T}_4 = 0$. As stated before, the infinite volume $T$-matrix can be reobtained, neglecting the left cut and tadpoles $L$-dependence, from Eq.~\eqref{eq:euluscher}. 

\subsection{Fixing the free parameters}
\label{subsec:fits}

The partial wave amplitudes, calculated in the $N/D_A$ method and in the IAM, depend on four LECs, $\bar{l}_i = 1,\ldots,4$, and, in the case of the $N/D_A$ method, also on the subtraction constant $a(\mu)$ and the parameters related to the $\rho$ bare field, $M_\rho$ and $G_V$. These parameters are fixed by reproducing scattering data as well as lattice results. For $I=0$, the phase shifts that we fit contain the very precise data of $K_{e4}$ decays below $\sqrt{s} = 400\ \textrm{MeV}$ \cite{Rosselet:1976pu, Pislak:2001bf, Pislak:2003sv, Batley:2007zz, Batley:2010zza}. Above that energy, the data of Ref.~\cite{Froggatt:1977hu} and the average of different experiments \cite{ochsthesis, Grayer:1972:NS, Estabrooks:1973zd, Kaminski:1996da, Hyams:1973zf, Protopopescu:1973sh}, as used \textit{e.g.} in Ref.~\cite{Oller:1998zr}, are taken into account. For $I=1$ and $J=1$, we use the data of Refs.~\cite{Estabrooks:1974vu, Lindenbaum:1991tq}. For $I=2$, the data come from Refs.~\cite{Losty:1973et,Hoogland:1977kt}. The lattice QCD results of Refs.~\cite{Noaki:2008gx,Beane:2007xs,Feng:2009ij} on the pion mass dependence of $f_\pi$ and $a_0^2$, the $I=2$ $J=0$ scattering length, are also fitted. 

\begin{figure}[ht]\centering
\includegraphics[width=1.0\linewidth]{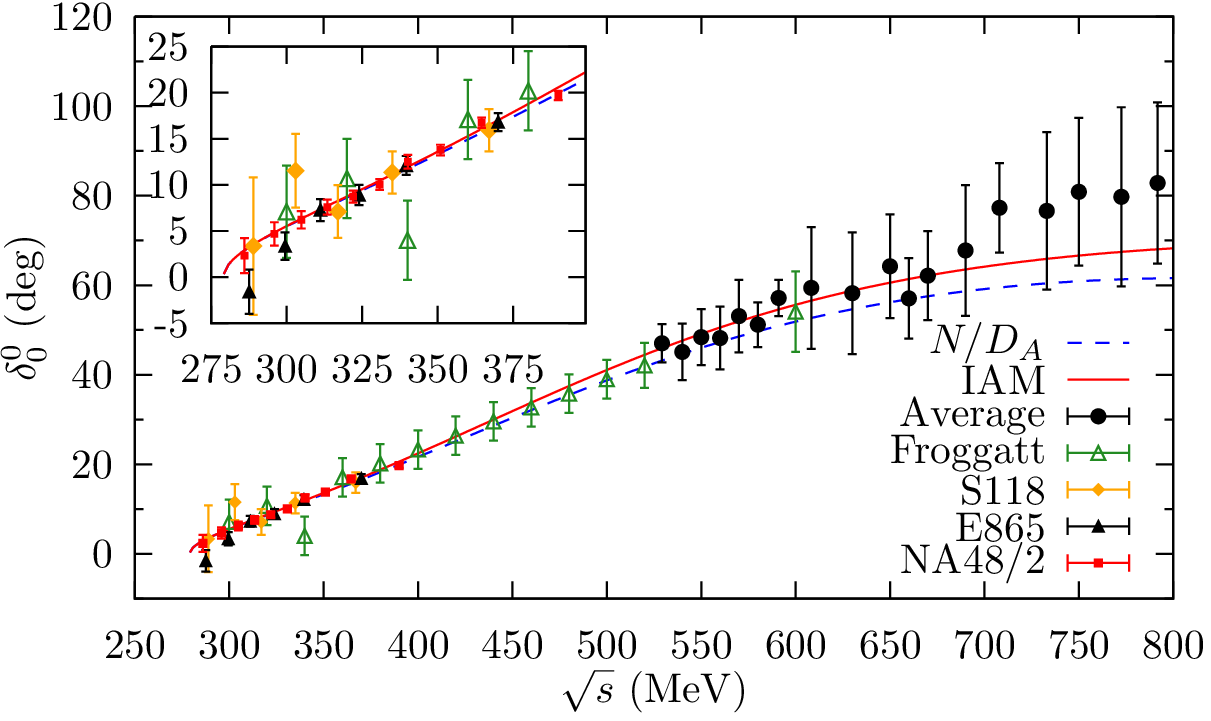}\\
\includegraphics[width=1.0\linewidth]{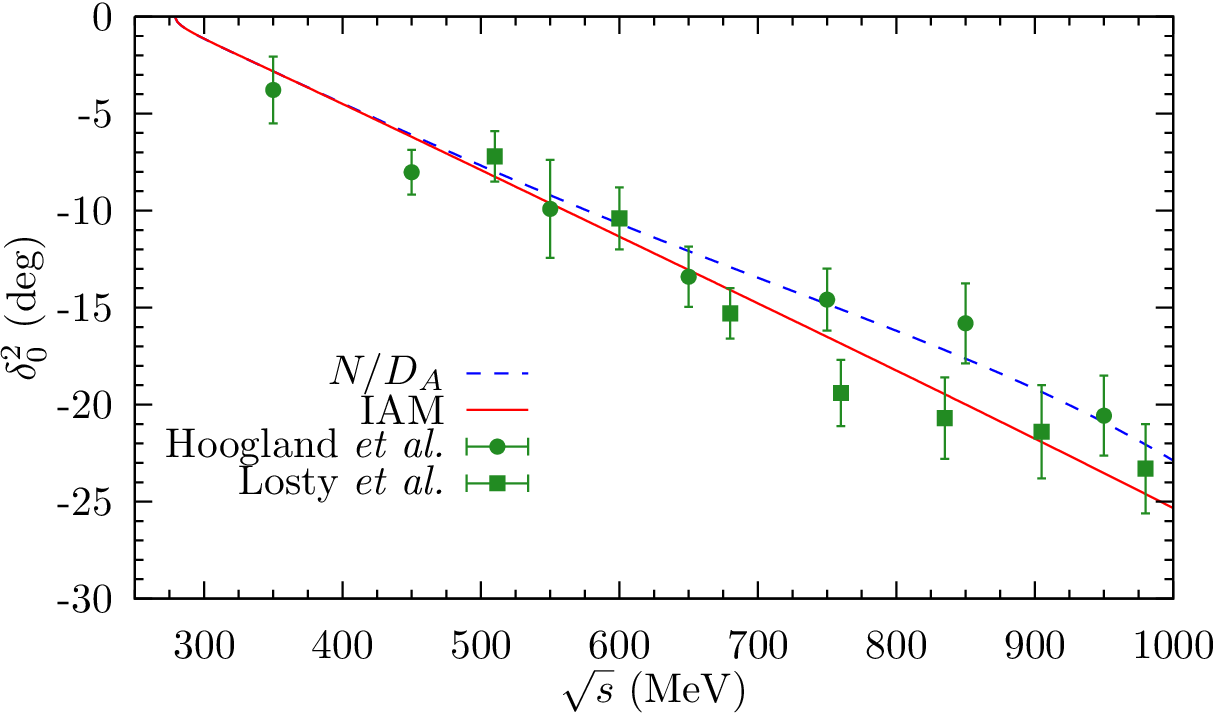}
\caption{(Color online) Comparison of our scalar $\pi\pi$ phase shifts to experimental data for $I=0$ (top panel) and $I=2$ (bottom panel). The (red) solid line 
 is our fit with the IAM, whereas the (blue) dashed one corresponds to the fit with the $N/D_A$ method. The inset in the top panel shows in more detail the low energy $K_{e4}$ decays data. The data for $I=0$ are from the $K_{e4}$ decay data of Refs.~\cite{Rosselet:1976pu, Pislak:2001bf, Pislak:2003sv, Batley:2007zz, Batley:2010zza}  and other data from Refs.~\cite{Froggatt:1977hu, ochsthesis, Grayer:1972:NS, Estabrooks:1973zd, Kaminski:1996da, Hyams:1973zf, Protopopescu:1973sh}. For  $I=2$ the phase shifts are from Refs.~\cite{Losty:1973et, Hoogland:1977kt}.\label{fig:fitph}}
\end{figure}

\begin{figure}[ht]\centering
\includegraphics[width=0.8\linewidth]{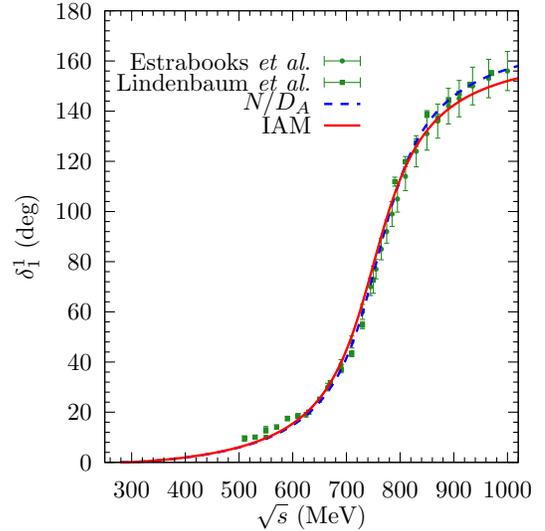}
\caption{(Color online) Comparison of our $I=1$ $J=1$ $\pi\pi$ phase shifts to experimental data. The (red) solid line shows our fit with the IAM, whereas the (blue) dashed one shows the fit with the $N/D_A$ method. The data are taken from Refs.~\cite{Estabrooks:1974vu, Lindenbaum:1991tq}.\label{fig:fitrho}}
\end{figure}

\begin{figure}[ht]\centering
\includegraphics[width=1.0\linewidth]{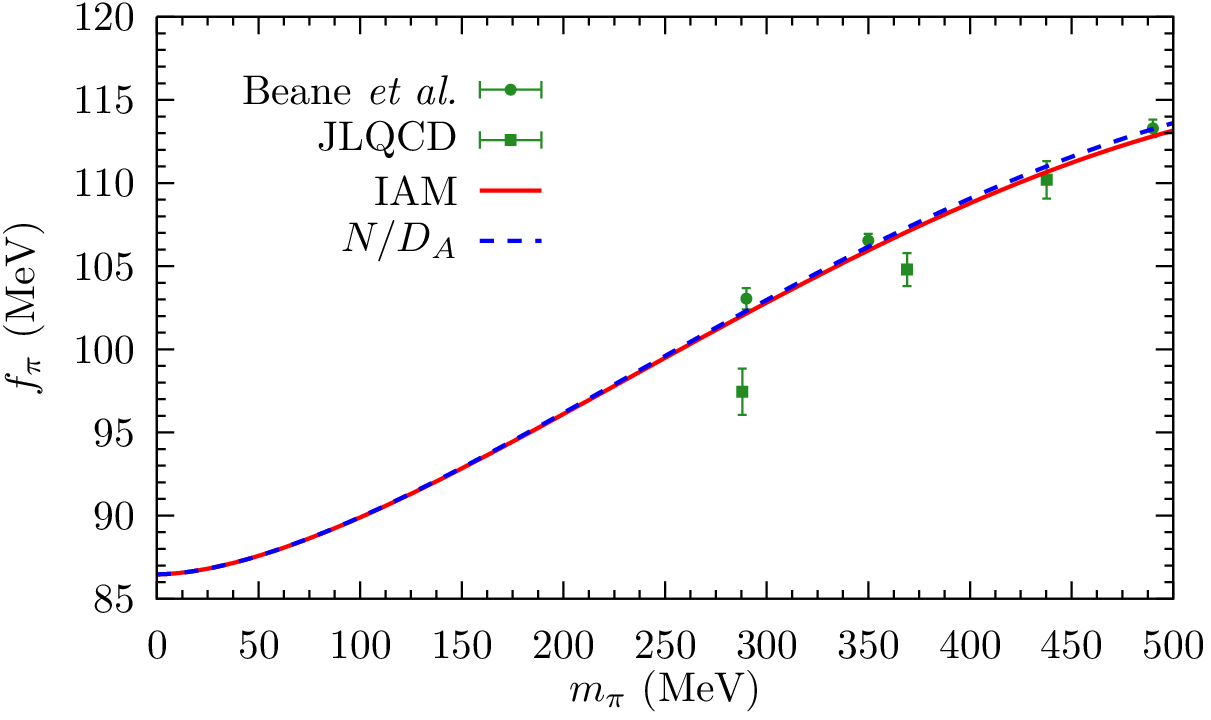}
\includegraphics[width=1.0\linewidth]{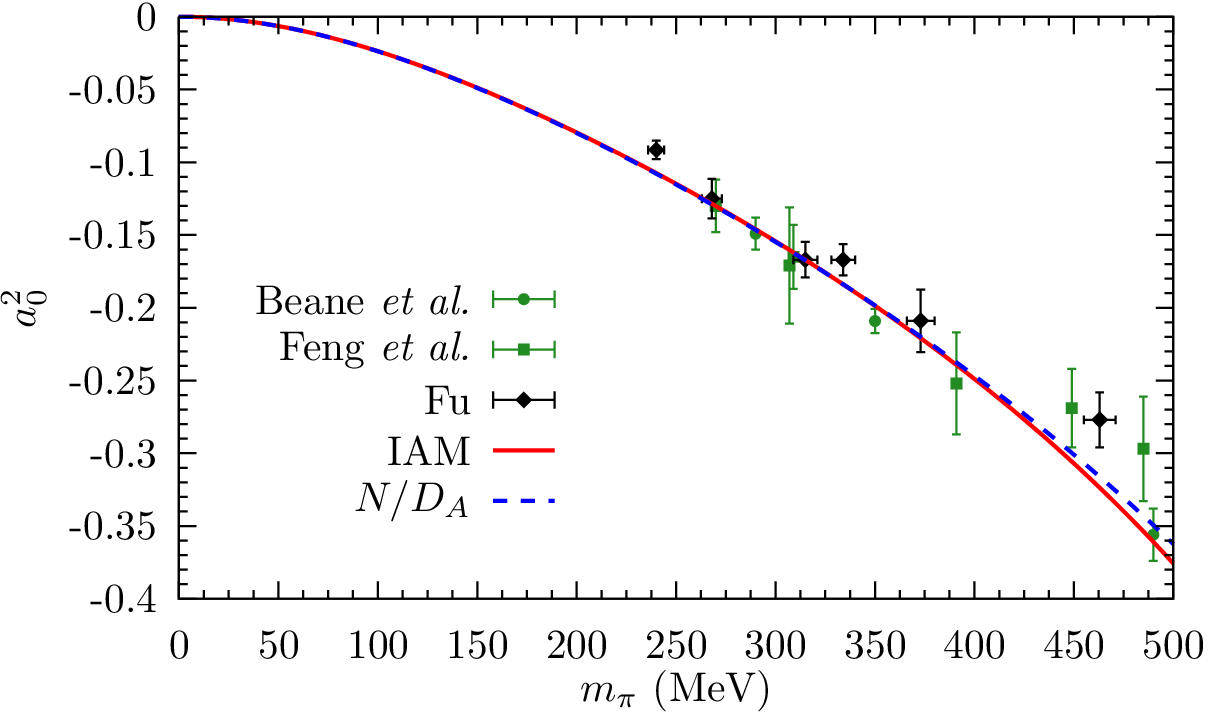}
\caption{(Color online) Dependence of $f_\pi$ (top panel) and $a_0^2$ (bottom panel) with $m_\pi$ as compared with lattice  QCD results. The (red) solid line is given by the fit with the IAM, whereas the (blue) dashed line represents that of the $N/D_A$ method. The data are from Refs.~\cite{Noaki:2008gx,Beane:2007xs,Feng:2009ij}. For comparison, we also show the recent results from Ref.~\cite{Fu:2013ffa}.
\label{fig:fitla}}
\end{figure}

We perform independent fits for the $N/D_A$ method and the IAM to the whole set of data, obtaining the parameters collected in Table~\ref{tab:params}. We note here the general agreement between the determinations of both methods. The LECs also compare well with the typical values given in the literature (see {\it e.g.} Ref.~\cite{Albaladejo:2012te} for comparisons and references). All the fitted data are shown in Figs.~\ref{fig:fitph}--\ref{fig:fitla}. The results of the $N/D_A$ are represented with dashed blue lines, whereas those of the IAM are shown with solid red lines. As can be see in these figures both models are compatible within the experimental uncertainties. Of course, the main aim in the present work is the relative change when  passing from infinite to finite volume and not the actual values of the phase shifts at infinite volume. Nevertheless, one should 
take as starting point an approach that can reproduce physical data in a fair way, as this is the case for both the N/D and IAM unitarization 
methods. 
  Finally, for completeness, in Table~\ref{tab:masses} we also give the values for the predicted pole positions and couplings of the $\sigma$ and $\rho$ mesons for both methods, that are compatible within the typical errors for these parameters (see PDG, Ref.~\cite{Beringer:1900zz}, and references therein related to the $\sigma$ and $\rho$ mesons).


\begin{table}[ht]\begin{center}\begin{tabular}{ccccccccc} \toprule\toprule
Fit & $\bar{l}_1$ & $\bar{l}_2$ & $\bar{l}_3$ & $\bar{l}_4$ & $a(\mu)$ & $M_\rho\ \text{(MeV)}$ & $G_V\ \text{(MeV)}$ \\ \bottomrule
$N/D_A$ & $-1.7$ & $5.9$ & $4.7$ & $4.0$ & $-1.5$ & $811$ & $62$ \\
IAM   & $-0.4$ & $5.6$ & $5.5$ & $4.0$ & - & - & - \\ \toprule\toprule
\end{tabular}\end{center}
\caption{Values of the parameters obtained from the fit of the data for the $N/D_A$ and the IAM methods.\label{tab:params}}
\end{table}
\begin{table}[h!]\begin{center}
\begin{tabular}{ccccc} \toprule\toprule
Method & $\sqrt{s_\rho}\ \text{(MeV)}$ & $|g_\rho|\ \text{(GeV)}$ & $ \sqrt{s_\sigma}\ \text{(MeV)}$ & $|g_\sigma|\ \text{(GeV)}$ \\ \bottomrule
$N/D_A$ & $758 - 70i$ & $2.4$ & $434 - 251i$ & $3.0$ \\
IAM   & $750 - 74i$ & $2.5$ & $439 - 236i$ & $2.9$ \\ \toprule \toprule
\end{tabular}\end{center}
\caption{Predicted values of the $\rho$ and $\sigma$ pole positions and 
couplings  for the $N/D_A$ and the IAM methods.
The couplings are defined, for both the $S$-- and $P$--wave, as $g^2_R=\lim_{s\to s_R}(s-s_R)T(s)$.\label{tab:masses}}
\end{table}

\section{Results}
\label{sec3}
\begin{figure}[ht]
\centering
\includegraphics[width=1.0\linewidth]{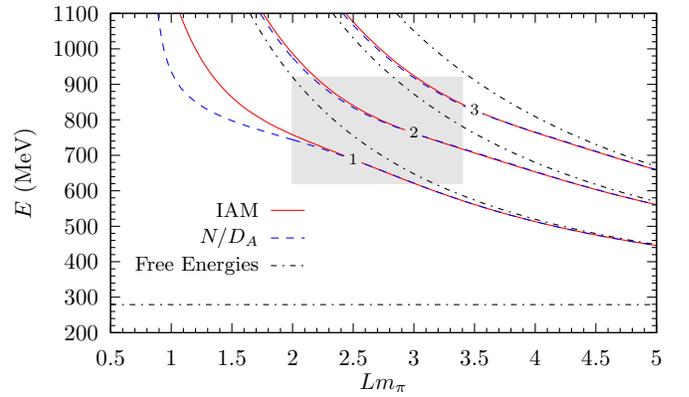}
\caption{(Color online) First three energy levels above threshold for the $\pi\pi$ interaction in the $\rho$ channel. The blue dashed lines correspond to the $N/D_A$ method, and the red solid ones stand for the IAM ones. The black dot-dashed curves correspond to the non-interacting energies.\label{fig:5}}
\end{figure}
\begin{figure}[ht]
\centering
\includegraphics[width=0.85\linewidth]{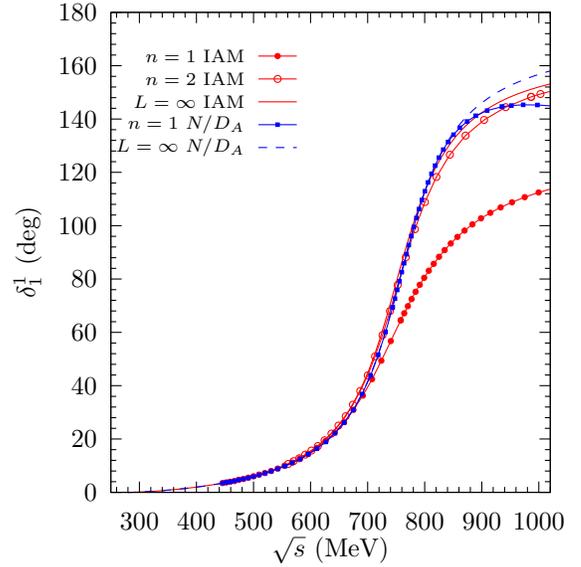}
\caption{(Color online) Phase shifts for the $\rho$-channel $\pi\pi$ scattering. The lines with closed (open) circles represent the results obtained with the first (second) level of the IAM. The lines with squares stand for the results of the first level of the $N/D_A$ method. The infinite volume results (that is, the ones in Fig.~\ref{fig:fitrho}) are also shown for comparison. The solid line corresponds to the phase shifts of the IAM, whereas the dashed one gives the results of the $N/D_A$ method.\label{fig:6}}
\end{figure}

In Fig.~\ref{fig:5} we show  the first three energy levels for different values of the cubic box size, $L$, obtained from the poles of the $I=1$, $J=1$ scattering amplitudes in the finite box as explained in the previous sections. 
The (red) solid lines stand for the IAM results, whereas the (blue) dashed ones represent the outcome of the $N/D_A$ method. The dot-dashed lines represent the free $\pi\pi$ energies in the box induced by  the periodic boundary conditions, which correspond to $E=2\sqrt{(\frac{2\pi}{L})^2 n+m_\pi^2}$, $n \in \mathbb{N}$. The meaning of the shadowed box will be explained below. 
We can see that the differences between the $N/D_A$ and IAM methods
are only relevant for the first level and for box sizes about
$Lm_\pi<2$.

If some points on the $E$ vs. $L$ plots are provided, for instance by 
an actual lattice QCD calculation, 
the scattering amplitudes (or related magnitudes like
phase shifts) can be obtained in the physical infinite volume case. 
This procedure is usually called the ``inverse problem'' and it 
is the final goal of the L\"uscher method or the method of Ref.~\cite{misha} 
(see Eq.~\eqref{eq:euluscher}). In Fig.~\ref{fig:6} 
we show the $I=1$, $J=1$ phase shifts obtained for the different methods
by solving the ``inverse problem'' from the first two energy levels of
Fig.~\ref{fig:5}. For comparison, we also show the phase shifts obtained with both methods in the infinite volume case (already  shown in Fig.~\ref{fig:fitrho}). 

For the results obtained with the level 2 the difference for both
methods are small and compatible with the experimental uncertainties of
the phase shifts. For the $N/D_A$ method, indeed, the phase shifts 
calculated from level 2 and in the infinite volume case are so similar that we only show the latter one. This means that for 
 points coming from the level 2 (or higher) the $L$--dependence of the potential has a minor effect in the
resolution of the ``inverse problem''.
However, if one uses points from level 1
 the differences are more important for
energies above $700\mev$. This effect is mild for the $N/D_A$, while it is rather strong in the case of the IAM. 
The reason for this different behavior between both methods
is that in the $N/D_A$ the $\rho$ meson is included as an explicit pole with
fixed bare mass. The physical $\rho$ generates  from the dressing of this bare field due to $\pi\pi$ rescattering.
 Therefore,  for energies close to the (bare) $\rho$ mass this
$L$-independent 
term dominates the amplitude. This is unlike the IAM
case where the $\rho$ resonance is generated from information codified in
the LECs and hence gets much more affected by the $L$
dependence of the internal dynamics that generates the amplitude.

These results are very illustrative and can be used to select
the range of values of $L$ where the finite box dependence
of the potential can be safely neglected to obtain the parameters of the
$\rho(770)$ meson, or phase shifts in this channel. This is represented by the shadowed region in
Fig.~\ref{fig:5}. The horizontal bounds represent the range $m_\rho\pm
\Gamma_\rho$ which would be desired in order to get the resonant
shape of the $\rho$. The left vertical bound represents the limit where
the results from level 1 gives acceptable results for both methods. This
corresponds approximatly to energies below $720\mev$ and  to
an $L$ value of about $2m_\pi^{-1}$. This means that using lattice
points from level 1, one can only obtain the same results for both
methods for the $\rho$ phase shifts for
energies below $720\mev$. In order to get similar results for both
methods for the right
part of the $\rho$ resonance  shape one must consider points in the second
level for $L$ values between $2-3m_\pi^{-1}$.

The $N/D_A$ and IAM models considered in the present work are two
 sound  models rooted in basic properties of strong interactions as unitarity, 
analyticity and chiral symmetry and have been rigorously and thoroughly tested in 
many processes in infinite volume. By considering the $L$--dependence that stems 
from such models we conclude with some confidence 
  that if one uses points obtained from the second level 
of Fig.~\ref{fig:5} in order to extract the $\rho$ meson, then it is safe to
neglect such $L$--dependence in the potential and proceed via standard
L\"uscher method or its version of \cite{misha} to solve the inverse problem. 
If points are used from lattice data
from level 1 then the $L$--dependence can certainly be neglected to
generate the low part of the $\rho$ resonance but we cannot make the same claim for the upper part of the resonance due to the strong $L$
dependence that we obtain for the IAM method. 

\begin{figure}[h!]\centering
\includegraphics[height=8.0cm,keepaspectratio]{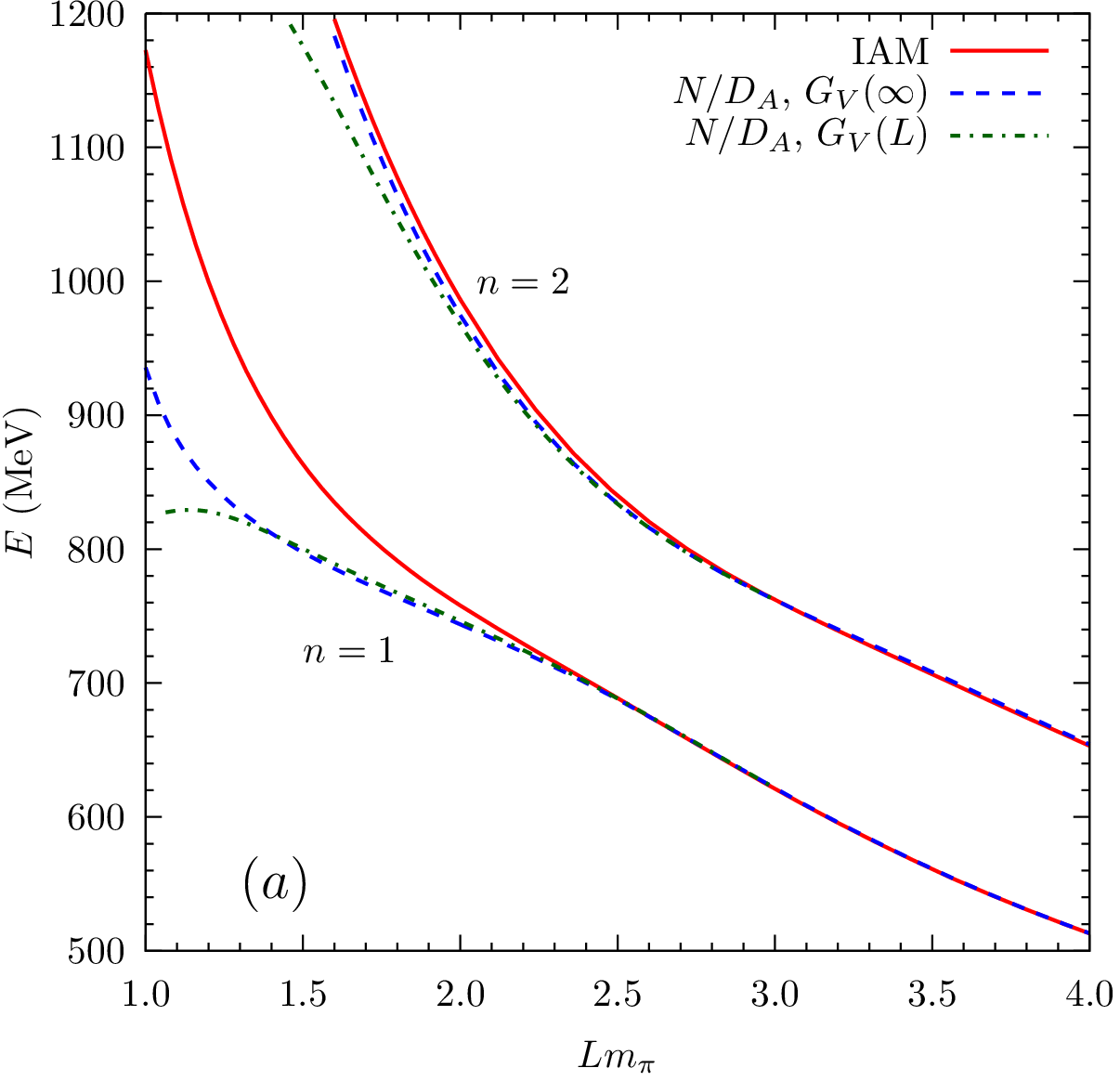}
\includegraphics[height=8.0cm,keepaspectratio]{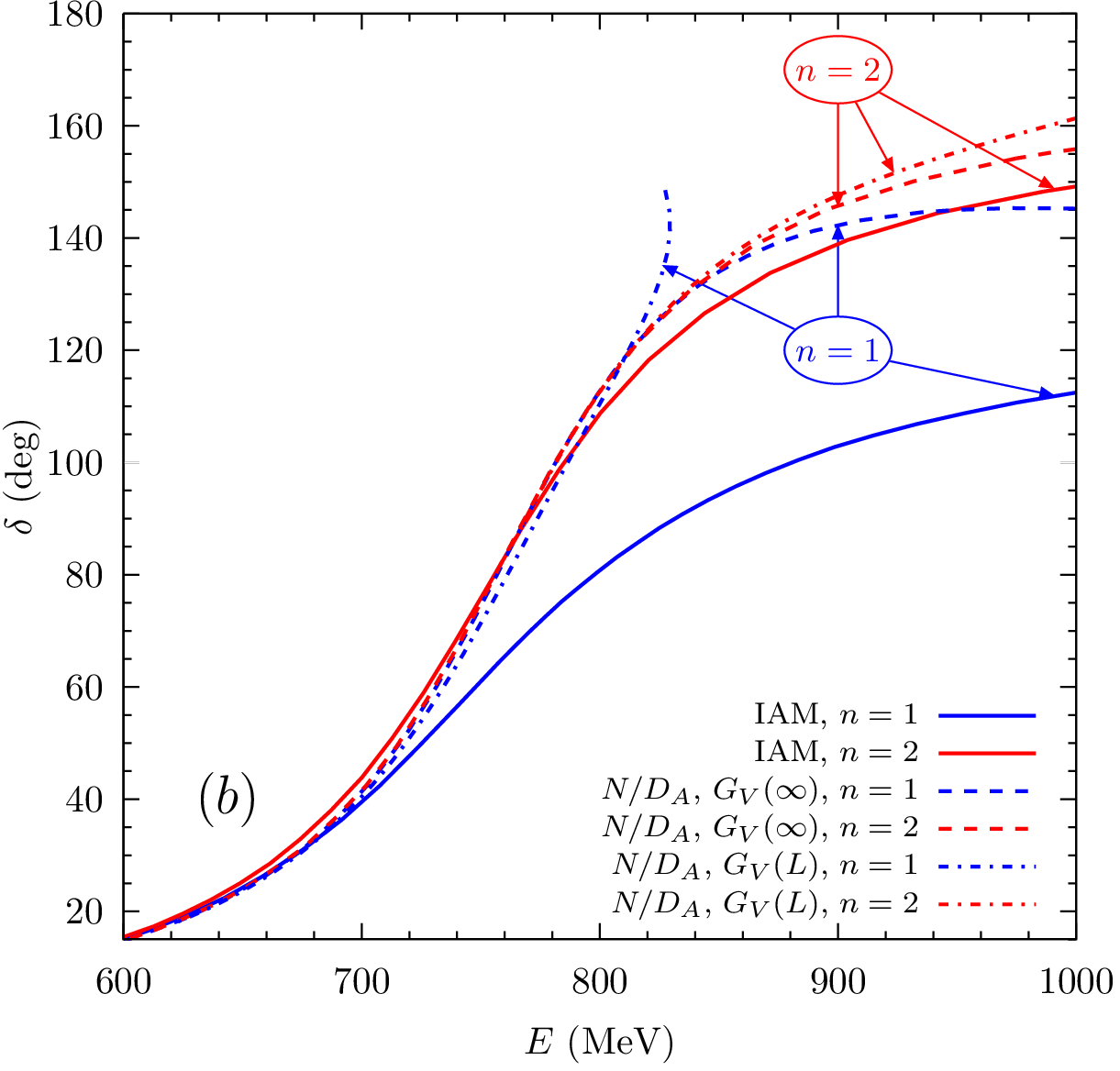}
\caption{(Color online) Comparison between the results obtained with the original $N/D_A$ and IAM methods and the results obtained with the additional volume dependence of $G_V$, Eq.~\eqref{eq:NewGV}. (a) Energy levels calculated with the original $N/D_A$ and IAM (blue dashed and red solid lines, respectively) are compared with those obtained including the volume dependence of $G_V$ (green dot-dashed lines). (b) Phase shifts calculated with the original $N/D_A$ and IAM (solid and dashed lines, respectively), as compared with those obtained considering $G_V(L)$ (dot-dashed lines). For each case, the blue (red) lines are obtained with the $n=1$ ($n=2$) energy level.\label{fig:GVL}}
\end{figure}

We now elaborate on the possible $L$ dependence of the $\rho$ coupling $G_V$. The original type of Kawarabayashi-Suzuki-Riazuddin-Fayyazuddin (KSRF) relation 
\cite{115g} predicts $G_V=f_\pi/\sqrt{2}$. This relation can also be derived from the high-energy constraint of the pion vector form factor 
at tree level \cite{Ecker:1989yg}. However, when the latter study is extended up to the one-loop level \cite{112g} it gives rise to 
the relation $G_V=f_\pi/\sqrt{3}$ that it is considered to be valid in the large $N_C$. 
  This modified KSRF-like relation was also confirmed in other contexts: $\pi\pi$ scattering \cite{1g,46g}, radiative $\tau$ decays \cite{113g}, 
extra-dimension model for $\pi\pi$ scattering \cite{116g} and within spectral-function sum rules and semi-local duality \cite{go}. We then consider quite reasonable to take that $G_V$ is proportional to $f_\pi$ and, from the $L$ dependence of the latter, to obtain that of $G_V$. This extra 
$L$ dependence is now taken into account and reads
\begin{equation}\label{eq:NewGV}
G_V(L) = G_V(\infty) \frac{\displaystyle 1 - \frac{\tilde{H}^D(L)}{f_\pi^2} + \frac{m_\pi^2}{16\pi^2 f_\pi^2} \bar{l}_4}{\displaystyle 1 + \frac{m_\pi^2}{16\pi^2 f_\pi^2} \bar{l}_4}.
\end{equation}
One comment is in order here. We should note that the amplitudes involving $\rho$-exchange are already $\mathcal{O}(p^4)$, and hence the above modification introduces terms of ${\cal O}(p^6)$ and higher in the evaluation of $V(s)$ in Eq.~\eqref{Tinv}. However, as argued above, it is of interest 
in order to identify sources of $L$ dependence in the resulting scattering amplitudes by using the $N/D_A$ method. Next we discuss the results that stem by considering this additional volume dependence of the amplitude.
 The energy levels and phase shifts obtained now are represented in Fig.~\ref{fig:GVL}. The upper panel,  Fig.~\ref{fig:GVL}(a), shows 
by the (green) dot-dashed lines the the new energy levels obtained.
 For comparison, we also display the original $N/D_A$ and IAM results, already shown in Fig.~\ref{fig:5}, by the 
 (blue) dashed and (red) solid lines, respectively. One can observe that the new energy levels are very similar to the previous ones
 for $Lm_\pi \geqslant 1.7$. However, the new $n=1$ energy level starts to be different from the original $N/D_A$ one 
for $Lm_\pi \leqslant 1.5$. Indeed, for smaller values of $L$ the novel $n=1$ energy level even decreases for decreasing values of $L$ and 
tends to values  near the bare $\rho$ mass ($M_{\rho} = 811\ \text{MeV}$). This is due to the fact that $G_V$ becomes very small
 for these small values of $L$ (for example, $G_V(L=1.5m_\pi^{-1})$ is around $20\ \text{MeV}$, whereas $G_V(\infty)=62\ \text{MeV}$). 
Thus, the effect of the bare $\rho$ meson can only be felt for energy values very close to the bare $\rho$ mass.

 Although the $\mathcal{O}(p^6)$ contribution to $V(s)$ considered here is only partial, 
it may indicate that the approach is not valid for values of $Lm_\pi$ in the range $1.0-1.5$.
 Let us recall here that these $L$ values are far from the range $Lm_\pi \geqslant 2$, confirming our previous
 conclusion that  this lower bound can be taken as safe in order to extract reliable $\pi\pi$ phase shifts and  information about the $\rho$ meson. 
In the bottom panel, Fig.~\ref{fig:GVL}(b), we show by the dot-dashed lines the phase shifts obtained with the new energy levels, 
together with the original ones, already shown in Fig.~\ref{fig:6}, resulting from the IAM (solid lines) and the $N/D_A$ method (dashed lines). 
For each of these three models, the phase shifts obtained with the energy levels $n=1$ and $n=2$ are shown by the blue and red lines, respectively.
 As expected,  the three approaches are rather similar for the $n=2$ level. The differences between the phase shifts calculated
 from the $n=1$ energy level obtained with the two $N/D_A$ results (with and without volume dependence of $G_V$) are large for energy values 
around $E=820\ \text{MeV}$. The deviation of the new calculated phase shifts with respect to the original ones (which were already close
 to the infinite volume phase shifts) for this range of energies (corresponding to values of $Lm_\pi$ in the range $1.0-1.5$)
 confirms the previous discussion on the large effects that stem from this energy range with low $L$  and that invalidates the use of 
Eq.~\eqref{eq:euluscher}. On the other hand, the agreement between the phase shifts calculated 
with the two versions of the $N/D_A$ method (with and without volume dependence of the coupling $G_V$) 
 supports the results obtained with energy levels that involve values of the volume $Lm_\pi \geqslant 2$.
\vspace{0.5cm}

\section{Summary}\label{sec:summary}

We have made a study of the $\pi\pi$ scattering in the $\rho(770)$ channel in a
finite box. In particular, we have studied the effect of the exponentially
suppressed $L$-dependence coming from tadpoles and crossed channel loops that
is usually neglected when extracting physical quantities from lattice results 
in a finite volume via L\"uscher approach or the method of Ref.~\cite{misha}. 
To do so we have used two realistic models to describe
$\pi\pi$ scattering, the $N/D_A$ method and the Inverse Amplitude Method (IAM).
These approaches describe fairly well the scattering and the $\rho(770)$ resonance in
infinite volume and, when modified to describe the $\pi\pi$ energy levels in a finite box,
they also incorporate the suppressed $L$-dependence mentioned above. We obtain for each
method the $\pi\pi$ energy levels in a finite box as a function of the box size $L$, and
study when the L\"ushcer approach or the method of Ref.~\cite{misha} can be safely used
to obtain the infinite volume phase shifts around the $\rho(770)$ region. We conclude
that if one uses lattice sizes $L>2 m_\pi^{-1}$, the exponentially suppressed $L$--dependence
is in fact numerically negligible, and the same results are obtained from the 
levels derived from either the $N/D_A$ method or the IAM, 
which also agree with the infinite volume phase shifts. 
However, if the high energy part of the
$\rho(770)$ shape is to be reproduced from the first energy level,  one needs to
use $L$ values smaller than $2m_\pi^{-1}$. Then, the results from the $N/D_A$ and IAM 
levels are quite different, which suggests that the neglected $L$--dependence in this case
might be numerically important also in lattice calculations, 
and the phase shifts obtained inaccurate.

\label{sec4}

\section*{Acknowledgments}
We thank Eulogio Oset for prompting us to perform this research. This work is partly supported by DGICYT contracts  FIS2006-03438,
 the Generalitat Valenciana in the program Prometeo 2009/09, MINECO (Spain) and FEDER (EU) project Refs.~FPA2010-17806 and FIS2011-28853-C02-02, the Fundaci\'on S\'eneca  11871/PI/090, the EU Integrated Infrastructure Initiative Hadron Physics
Project under Grant Agreement n.~283286, and the DFG (CRC 16, ``Subnuclear Structure of Matter'' and CRC 110, ``Symmetries
and the Emergence of Structure in QCD'').

\end{document}